\documentclass[aps,prb,twocolumn,superscriptaddress]{revtex4-2}
\usepackage{natbib}
\usepackage{setspace}
\usepackage{bm}
\usepackage{amsmath}
\usepackage{graphicx}
\usepackage{xcolor}

\begin{document}

%\preprint{}

\title{A cold-source paradigm for steep-slope transistors based on van der Waals heterojunctions}


%\author{}
%\thanks{}
%\affiliation{}

\author{D. Logoteta}
\email{logotetad@gmail.com}
\altaffiliation{Universit\'e Paris-Saclay, Centre National de la Recherche Scientifique, Centre de Nanosciences et de Nanotechnologies, 91120 Palaiseau, France}
\affiliation{Dipartimento di Ingegneria dell'Informazione, Universit\`a di Pisa, Via G. Caruso 16, 56126 Pisa, Italy}

\author{J. Cao}
\affiliation{School of Electronic and Optical Engineering, Nanjing University of Science and Technology, Nanjing 210094, China}

\author{M. Pala}
\affiliation{Universit\'e Paris-Saclay, Centre National de la Recherche Scientifique, Centre de Nanosciences et de Nanotechnologies, 91120 Palaiseau, France}
\email{marco.pala@c2n.upsaclay.fr}

\author{P. Dollfus}
\affiliation{Universit\'e Paris-Saclay, Centre National de la Recherche Scientifique, Centre de Nanosciences et de Nanotechnologies, 91120 Palaiseau, France} 

\author{Y. Lee}
\affiliation{Integrated Systems Laboratory, ETH Z{\"u}rich, 8092 Z{\"u}rich, Switzerland}

\author{G. Iannaccone}
\affiliation{Dipartimento di Ingegneria dell'Informazione, Universit\`a di Pisa, Via G. Caruso 16, 56126 Pisa, Italy}


%\date{\today}

\begin{abstract}
The availability of transistors capable of operating at low supply voltage is essential to improve the key performance metric of computing circuits, \emph{i.e.}, the number of operations per unit energy.
In this paper, we propose a new device concept for energy-efficient, steep-slope transistors based on heterojunctions of 2D materials. 
We show that by injecting electrons from an isolated and weakly dispersive band into a strongly dispersive one, subthermionic subthreshold swings can be obtained, as a result of a cold-source effect and of a reduced thermalization of carriers. 
This mechanism is implemented by integrating in a MOSFET architecture two different monolayer materials coupled through a van der Waals heterojunction, combining the subthermionic behavior of tunnel field-effect transistors (FETs)  with the robustness of a MOSFET architecture against performance-degrading factors, such as traps, band tails and roughness. A further advantage with respect to tunnel FETs is that only an $n$-type or $p$-type doping is required to fabricate the device. In order to demonstrate the device concept and to discuss the underlying physics and the design options, we study through {\it ab-initio} and full-quantum transport simulations a possible implementation that exploits two recently reported 2D materials.
\end{abstract}

%\keywords{}

\maketitle

%\section{}
%\subsection{}
%\subsubsection{}

\section{Introduction}
In recent years, the number of operations per unit energy has emerged as the single most important performance metric for digital systems. This is true both in the domain of large data centers, that are essentially constrained by the total power consumed by the information technology equipment and by the backroom area (cooling plant, switchgears), and in the domain of distributed sensors for the Internet of things, that have extremely reduced energy budgets. 

Major improvements of performance per unit energy requires a change of paradigm with respect to the current MOSFET technology, in which the supply voltage is intrinsically lower-limited by the impossibility to decrease  the subthreshold swing (SS) of transistors below the thermionic limit of 60~mV/dec. This constraint is unavoidably associated to the thermal excitation of carriers within the high-energy Boltzmann tail of the electron distribution. 

Reducing the supply voltage is of the utmost importance: in a CMOS architecture, energy per logic gate switch scales with the square of the supply voltage, if only dynamic power consumption is considered, and super-quadratically, if also the static power consumption is taken into account.~\cite{Theis_In_2010}

Steep-slope transistors are a class of devices able to provide sub-thermionic SS and that, therefore, can be operated at lower supply voltages. Different physical mechanisms, including band-to-band tunneling (BTBT)~\cite{Seabaugh_IEEEproceedings_2010}, negative capacitance effects~\cite{salahuddin2008use}, impact ionization~\cite{gopalakrishnan2005}, and Mott transition\cite{nakano2012} have been explored as bases for the operation of these devices. 

In particular, tunnel FETs, in which the Boltzmann tail is filtered out {\it via} BTBT tunneling, have attracted much attention in the last decade, as devices with potentially sub-40 mV/dec SS, low control voltages and short delay times~\cite{Lu_JEDS_2014}. However, despite the encouraging theoretical predictions, the experimental observation of subthermionic SS has been proven extremely challenging. The switching process in tunnel FETs relies, indeed, on the modulation of the tunneling probability through a barrier, which is likely to be  dominated by disorder-induced nonidealities, as traps~\cite{Pala_TED2013}, band tails~\cite{Khayer2011Effects} and roughness~\cite{Conzatti_EDL2012}. 

As a way to circumvent these issues, in this paper we propose an alternative design paradigm that exploits the tailoring opportunities offered by the 2D materials platform. Within this approach, the subthermionic capabilities of the devices are still driven by an energy filtering effect, as in tunnel FETs, but the switching is obtained by modulating the height of a thermionic barrier, as in MOSFETs. According to previous studies~\cite{Esseni_TED2013}, the operation of these devices is expected to be significantly less sensitive to the performance-degrading factors plaguing the tunnel FETs. As a further advantage, only an $n$-type ($p$-type) doping is needed for $n$ ($p$) devices, while a combination of $n$-type and $p$-type is always needed in tunnel FETs. We demonstrate this paradigm by numerically investigating the operation and the potential of a representative device, based on two recently reported 2D materials: the ternary KTlCl$_4$  monolayer and the Au$_2$S monolayer in its $\alpha$ phase.

\section{Device concept}

\begin{figure}
    \centering
	\includegraphics[width=0.8\columnwidth]{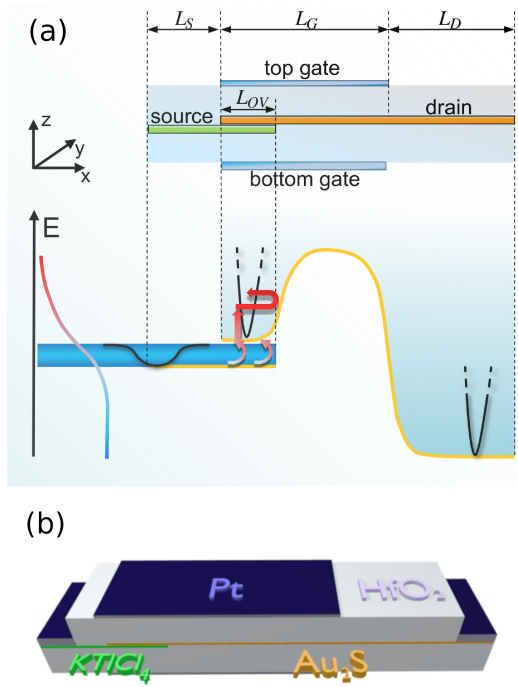}
	\caption{\protect \footnotesize (a) Sketch of the transistor structure (top) and of the corresponding band diagram (bottom) when the device is not conducting. Electrons are injected in the narrow band of the source monolayer, leaving essentially empty the high-energy Boltzman tail. Next, electrons tunnel through a van der Waals heterojunction in the strongly dispersive monolayer implementing the channel and the drain, where they undergo a partial thermalization before being reflected back by the barrier (red arrows). (b) Sketch of the representative device studied in the paper, in which the source and channel/drain regions are implemented by monolayers of KTlCl$_4$ and Au$_2$S, respectively.}
  \label{fgr:concept}
\end{figure}

%\begin{figure}[b]
%\includegraphics{FIGURES/Fig1.eps}% Here is how to import EPS art
%\caption{\label{fgr:concept} A figure caption. The figure captions are
%automatically numbered.}
%\end{figure}

A sketch of the proposed architecture is illustrated in Fig.~\ref{fgr:concept} (a). The device consists of two monolayers, one of which acts as source, while the other implements the channel and the drain. The monolayers overlap over a length $L_{\rm OV}$, which establishes a connection between them {\it via} a van der Waals heterojunction~\cite{Na_ACS2019, Iannaccone2018}. The current is controlled by a top and a bottom gate of the same length $L_{\rm G}$, separated from the 2D materials by two insulating layers. The source material and the drain are both $n$- or $p$-doped, according to the required device polarity. Overall, this architecture differs from that of a monolayer-based double-gate MOSFET only by the presence of a source injector coupled to the channel. 

In order to explain the device operation, a sketch of the band diagram of an $n$-type transistor in the non-conducting regime is also dislayed in Fig.~\ref{fgr:concept} (a). 
The source injector is designed to provide an energy-filtering effect able to cut off the thermal tail of the Boltzmann distribution of carriers (a so-called ``cold-source'' effect~ \cite{Qiu2018,Logoteta2019,Marin2020,Liu_2020}). This effect can be obtained by exploiting isolated narrow bands, intrinsically present in the band structure of several 2D materials \cite{mounet2018two}. When the transport occurs in one of these bands, the electron energy is confined to a small window, and the Boltzmann tail remains essentially empty of carriers (see Fig.~\ref{fgr:concept} (a)). 
%We remark that the narrowness of these bands is unavoidably associated to large effective masses and high density of states, which promotes the electron-phonon scattering. 

By tunneling through the van der Waals heterojunction, electrons can reach the other monolayer, where they can undergo a partial thermalization driven by phonon absorption. However, if the electron-phonon scattering rates in this material are small enough, only a small percentage of electrons can acquire enough energy to overcome the potential barrier, and most of them will be reflected back (see Fig.~\ref{fgr:concept} (a)). Within this assumption, the device basically works as a traditional FET with a reduced leakage thermionic current in the subthreshold regime, and can thus exhibit subthermionic capabilities.   
Suitable materials to implement the channel and drain layer should thus have high phonon-limited mobility, usually associated to very small effective masses (and, therefore, low density of states)~\cite{LundstromBook}. The transport in these materials occurs within strongly dispersive bands, the opposite of what is required in the source injector, where the high effective mass and density of states promote a strong electron-phonon scattering. 
%If these requirements on the materials are fulfilled, the device basically works as a traditional FET with a reduced leakage thermionic current in the subthreshold regime, and can thus exhibit subthermionic capabilities. 

A suitable band alignment at the heterojunction is of critical importance to ensure good performance. Ideally, the conduction bands of the monolayers should be close enough or even overlapping, to favor the interlayer tunneling and ensure good current levels in the ON state. In this respect, we notice that extending the top and bottom gates over the heterojunction, as shown in Fig.~\ref{fgr:concept} (a), provides a way to tune the band offset as a function of the difference between the work functions of the metals used for the two gates.
%, and as a function of the gate voltage. 
Metal gates with different work functions (and held at the same voltage, as it is supposed here) induce an electric field that makes the potential on the two monolayers different. If such a field is strong enough, sizable rearrangements of the band alignment can be obtained, as already experimentally demonstrated (see \emph{e.g.} Ref.~\cite{Liu2016} and references quoted therein). 

%The change of the gate voltage, on the other hand, can influence the band offset based on the different density of states of the two materials. Due to the much smaller density of states, the potential in the channel material undergoes larger variations in response to a change in $V_{\rm GS}$ with respect to the potential in the source injector. Particularly, it is lowered more as $V_{\rm GS}$ increases, which results in a beneficial rapprochement of the conduction bands of the monolayers as the device approaches the full conduction regime. 

We remark that the tunnel barrier, intrinsically included in the device structure, is not demanded to play any role in the switching process. Accordingly, the impact on the subthreshold performance of defects like traps and roughness or disorder-induced band tails is expected to be much more limited with 
respect to tunnel FETs, and similar to that reported for MOSFETs~\cite{Esseni_TED2013}. If a proper design guarantees a sufficient transparency of the tunnel barrier, also the variability induced by such nonidealities, typically responsible for an increase of the tunneling probability, can be expected to be significantly smaller.  

%responsible for a detrimental increase of the tunneling probability in the OFF-state of tunnel FETs, and of significant degradation of the SS, in this case cannot sensibly alter the performance.

%The only possible contribution to the modulation of the current stems from the gradual adjustment of the band alignment in the overlap region as a function of the gate voltage (due to the markedly different density of states, the potential within the two monolayers undergoes different variations in response to a change in the gate voltage, entailing a change in relative position of the conduction band edges). However, this turns out to be a minor effect if a proper design guarantees a good transparency of the tunnel barrier.  

\section{Representative device model}

\begin{figure}
    \centering
	\includegraphics[width=0.9\columnwidth]{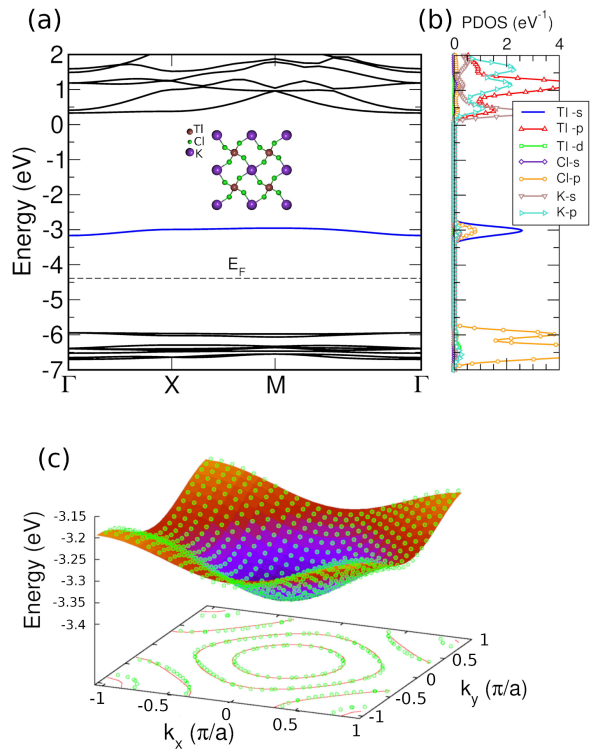}
	\caption{\protect \footnotesize	(a) Band structure of KTlCl$_4$ monolayer. The dashed line denotes the Fermi level. (b) Corresponding density of states projected over the valence atomic orbitals. (c) Plot over the first Brillouin zone of the isolated band highlighted in blue in (a), compared with the approximation obtained within the adopted tight-binding model (green circles).}
  \label{fgr:BS}
\end{figure}

In order to demonstrate in a practical case the proposed paradigm and to discuss more in detail the operation of this class of devices, we have numerically studied a possible implementation using two recently reported 2D materials that exhibit the required properties. 
The device is sketched in Fig.~\ref{fgr:concept} (b). 

We use as source material the ternary KTlCl$_4$, reported as a thermodynamically stable monolayer, easily exfoliable from its parent 3D compound~\cite{mounet2018two}. Its band structure is reported in Fig.~\ref{fgr:BS} (a), together with the projection of the density of states over the different atomic species (panel (b)). The isolated band highlighted in blue, with minimum in $\Gamma$ and approximately 0.2~eV wide, is used for the electron transport. This band exhibits a strong contribution from the squared sublattice of Tl atoms, and reflects the character of a simple tight-binding band deriving from an s-like orbital (see Fig.~\ref{fgr:BS} (b)). 
 
As channel and drain material, we choose the synthetic Au$_2$S in its $\alpha$ phase. This material exhibits a wide gap of 1 eV, which allows a good electrostatic control over the current, and a very small, isotropic electron effective mass of 0.06~$m_0$~\cite{wu2019two}. Its theoretical phonon-limited mobility is very high and has been estimated to be $8.45\times 10^4$ cm$^2$/V s at room temperature~\cite{wu2019two}. 

The transport in the device is addressed by means of full-quantum simulations within the non-equilibrium Green's function formalism. The electron-phonon scattering is included within a deformation potential approximation as detailed in the Method section.

%{\red Further, the transport properties of this device are addressed by using the non-equilibrium Green's function (NEGF) method and including electron-phonon scattering within the self-consistent Born approximation. In these calculations,}
In the transport simulation framework, the KTlCl$_4$ monolayer is described within a two-orbital tight-binding model. The Hamiltonian reads
$$
H(\bm k)=
\left(
\begin{matrix}
\epsilon_0+t_{\bm k} & t_{\bm k} \\
t_{\bm k}            & -\epsilon_0+t_{\bm k}
\end{matrix}
\right)
$$
where $t_{\bm k}=2\gamma[\cos(k_x a_x)+\cos(k_y a_y)]$ and $a_x=a_y=a_{{\rm KTlCl}_4}$  is the KTlCl$_4$ lattice constant. The lowest band of the model is used to describe the KTlCl$_4$ doubly spin-degenerate isolated band of interest. The parameters $\epsilon_0$ and $\gamma$ have been set to -0.5 eV and -0.025 eV, respectively, in order to fit the DFT band structure (see Fig.~\ref{fgr:BS} (c)). These values also guarantee that the artificial upper band of the tight-binding model is located at energies high enough to be empty of carriers and uninfluential on the transport. 

The Au$_2$S is described within an effective mass model, that turns out to be a good approximation, due to the large separation ($\approx$ 1.5~eV) of the conduction band minimum from the above-lying bands.  

While the simplicity of the band structures makes it easy to accurately model the two uncoupled materials, the particular values of the incommensurate lattice constants, $a_{{\rm KTlCl}_4}=6.577$ \AA\   and $a_{\rm Au_2S}=5.74$ \AA, hinders a description of the heterojunction directly based on {\it ab-initio} simulations. Such a description is commonly based on finding supercells for the top and bottom layer of approximately the same size and then apply a small amount of strain to make them commensurate~\cite{szabo_EDL2015,Cao2016operation}. Unfortunately, in the case at hand, the minimum size of the supercells compatible with reasonably low levels of strain is $\sim 7\times a_{{\rm KTlCl}_4}=4.6$ nm$^2$, and it is thus too large to allows a full {\it ab-initio} study. We are thus compelled to resort to a different, more approximate strategy. Accordingly, we modeled the coupled monolayers as effectively commensurate over a much smaller cell, of the same size as the KTlCl$_4$ unit cell, and we described the interlayer coupling by a spatially constant hopping amplitude $t_{\rm c}$. The value of $t_{\rm c}$ should be calibrated in such a way that the average interlayer tunneling probability per unit area is the same as in the real system. However, since the {\it ab-initio} simulation of the system of coupled monolayers is not affordable, the actual value of $t_{\rm c}$ cannot be accurately predicted, and the best we can do is to locate it within a reasonable range. In the following, we assume $t_{\rm c}=30$~meV. The choice of this particular value will be discussed at the end of the Results and Discussion Section.

%This parameter can be physically interpreted as a properly rescaled spatial average, over the above-%mentioned large supercell, of the total hopping amplitude between the atomic orbitals contributing to the %two bands considered in the model.  
%The scaling factor equals the ratio between the area of the KTlCl$_4$ unit cell and of the supercell, in %order to guarantee that the average interlayer coupling per unit area is the same as in the real system. 
%Accordingly, we modeled the interlayer coupling by an effective{\control{blue}, spatially constant} %hopping parameter $t_{\rm c}$ that we set at the value of 30 meV. The choice of this particular value %will be discussed at the end of the Results and Discussion Section. {\control{blue} $t_{\rm c}$} can be %physically interpreted as a spatial average over the above-mentioned supercell of the total hopping %amplitude between the atomic orbitals contributing to the two bands considered in the model.} 

In the same spirit, the band offset of the coupled monolayers is estimated by means of the Anderson's rule~\cite{Anderson1960}. The electron affinity of KTlCl$_4$ and Au$_2$S is computed to be 4.8~eV and 4.47~eV, respectively (see the Method section for the details), resulting in a gap of $\sim 0.1$~eV between the conduction bands of the coupled KTlCl$_4$-Au$_2$S system. However, as already mentioned, to enhance the interlayer coupling and prevent a significant degradation of the ON current, a condition in which the bands overlap is preferable. In order to achieve this goal, we choose for the top and bottom gates two metals with quite different work functions $\chi$. The top gate is made of Pt ($\chi=6.3$ eV), while the bottom gate is made of Ag ($\chi=4.7$ eV). The vertical electric field arising between the gates downshifts the potential in the Au$_2$S monolayer with respect to that in the KTlCl$_4$ one, enabling an overlap between the bands for all the regimes of operation of the device.

%The deformation potential of the dominant acoustic and optical phonon branches in KTlCl$_4$ and Au$_2$S were compued from first principles as detailed in the Methods section. {For the electron-acoustic phonon scattering, we use the deformation potential associated to the longitudinal acoustic modes provided in Ref.~\cite{wu2019two}. It turns out that the scattering with the transverse acoustic modes is negligible with respect to it, as expected from the near spherical symmetry of the conduction band. The coupling with flexural phonons is zero to the first order, due to the existence of a glide symmetry plane~\cite{Fischetti2016Mermin} (as the Au$_2$S is symmetric under glide reflections), and it is also neglected. The deformation potential associated to the optical phonons has been evaluated by fitting the (weakly energy-dependent and fairly isotropic) scattering rate of electrons in $\Gamma$ with the dominant polar optical mode.}

%We set the source and drain lengths of the simulated device to $L_{\rm S}\simeq 5$ nm and $L_{\rm D}\simeq 25$ nm, respectively. These values are large enough to guarantee the stabilization of the potential at the left and right end of the simulation domain. The much smaller value of the density of state in Au$_2$S with respect to KTlCl$_4$ requires  to fulfill this condition.

We set the source length of the simulated device to $L_{\rm S}\simeq 5$ nm. This value is large enough to guarantee the stabilization of the potential at the left end of the simulation domain. Due to the much smaller density of states in Au$_2$S with respect to KTlCl$_4$, fulfilling the same condition on the drain end requires a larger value for $L_{\rm D}$, which we set, accordingly, to $\simeq 25$ nm. We also assume that the monolayers are separated by the top and bottom gate by 3-nm thick layers of HfO$_2$ and that the KTlCl$_4$ monolayer and the Au$_2$S drain are $n$-doped with a concentration of $6\times 10^{13}$ cm$^{-2}$ and $4\times 10^{12}$ cm$^{-2}$, respectively. The source-to-drain bias ($V_{\rm DS}$) is set to 0.4~V. The OFF state of the transistor is defined by setting the value of the current in this state to $I_{\rm OFF}=10^{-4}$~$\mu$A/$\mu$m~\cite{IRDS}, while the ON state is obtained, as usual, at the gate voltage $V_{\rm GS}=V_{\rm GS}^{\rm off}+V_{\rm DS}$, where $V_{\rm GS}^{\rm off}$ is the gate voltage in the OFF state. 

The parasitic resistances associated to the source and drain contacts are neglected in our model. They are expected to mainly degrade the current close and above the conduction threshold, leaving essentially unaltered the subthreshold current, due to negligible voltage drop across contacts. We remark that, although the quest for an approach to obtain low-resistance contacts compatible with large scale manufacturability is still ongoing, several strategies have already been proposed and demonstrated, including one-dimensional edge contacts~\cite{Wang614}, van der Waals contacts~\cite{Wang2019} and surface roughness engineering~\cite{Banerjee2020}.

\section{Results and Discussion}

\begin{figure*}
    \centering
	\includegraphics[width=1.4\columnwidth]{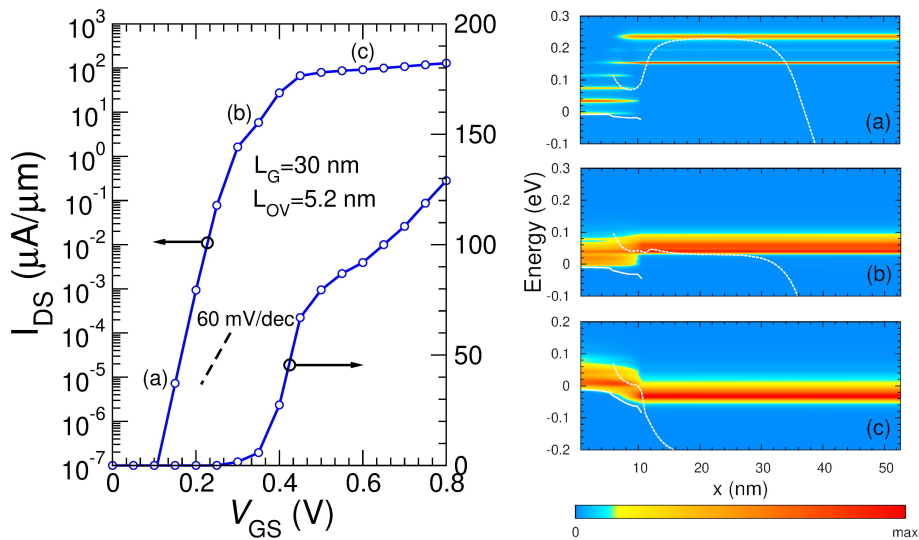}
	\caption{\protect \footnotesize	{\it Left}: transfer caracteristic of the KTlCl$_4$-Au$_2$S transistor for $L_{\rm G}=30$~nm and $L_{\rm OV}=5.2$~nm. $V_{DS}=0.4$~V. {\it Right}: Maps of the longitudinal current density spectrum along the device for (a) $V_{\rm GS}=0.15$, (b) $V_{\rm GS}=0.35$, and (c) $V_{\rm GS}=0.6$~V. The considered bias points are indicated by (a), (b) and (c) on the transfer characteristic. The white solid and dashed lines denote the bottom of the two conduction bands involved in the transport.}
  \label{fgr:IV_5.2nm_maps}
\end{figure*}

Fig.~\ref{fgr:IV_5.2nm_maps} illustrates the transfer characteristics of the device for $L_{\rm OV}=5.2$~nm and $L_{\rm G}=30$~nm. The results confirm the subthermionic capabilities of the transistor, which is able to achieve a minimum subthreshold swing of 26~mV/dec. The ON current amounts to $I_{\rm ON}\approx 90$ $\mu$A/$\mu$m, corresponding to an $I_{ON}/I_{OFF}$ ratio of around $10^6$. 

In order to gain more insights into the operation of the device, in Fig.~\ref{fgr:IV_5.2nm_maps} it is also reported the spectrum of the longitudinal current for three different values of $V_{\rm GS}$. For $V_{\rm GS}$ close to the OFF state (panel (a)), electrons are mostly injected from the source contact close to the bottom of the KTlCl$_4$ conduction band and rised at energies above the top of the barrier by the strong phonon absorption. This is in sharp contrast with the behavior usually observed in traditional MOSFETs, in which the injection of carriers directly occurs at energies closer to the top of the channel barrier. Significant spectral components at energies under the top of the barrier, related to source-to-drain tunneling processes, are also present. Both the strength of the electron-phonon interaction and the tunneling probability through the channel barrier are therefore important to determine the leakage current in the OFF state.   

As $V_{\rm GS}$ increases and the channel barrier is pushed down (panel (b)), optical phonon absorption gradually decreases and concentrates at the right edge of the heterojunction, where the electrons propagating in KTlCl$_4$ face the monolayer boundary. This trend is expressed by a step-like feature in the current spectrum near $x=10$~nm. An analogous behavior can be observed close to the ON state (panel(c)), where, however, the transport is dominated by phonon emission processes and the sign of the step is reversed. All the maps indicate that the electron thermalization outside the source and the overlap region is negligible over the considered scale lengths.

\begin{figure}
    \centering
	\includegraphics[width=0.8\columnwidth]{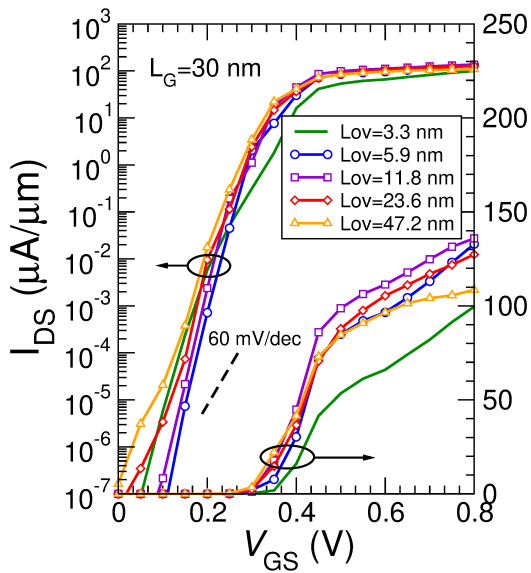}
	\caption{\protect \footnotesize	Transfer characteristics of the transistor for $L_{\rm G}=30$~nm and different values of $L_{\rm OV}$. $V_{DS}=0.4$~V.}
  \label{fgr:IV_LOV}
\end{figure}

\begin{figure*}
    \centering
	\includegraphics[width=1.7\columnwidth]{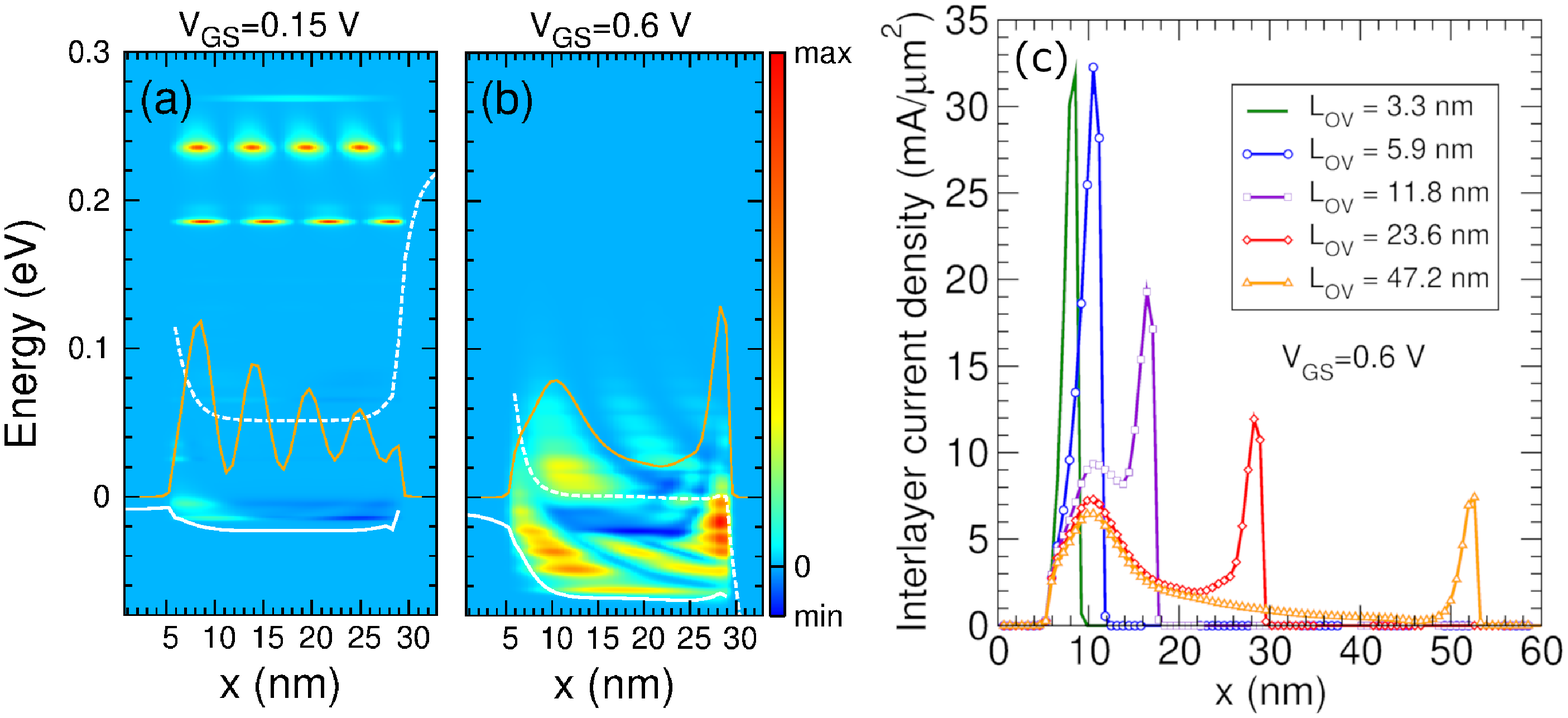}
	\caption{\protect \footnotesize	Maps of the interlayer current spectrum for $L_{\rm OV}=23.6$~nm and (a) $V_{\rm GS}=0.15$~V and (b) $V_{\rm GS}=0.6$~V. The orange curves superimposed to the maps are the plot (in arbitrary units) of the interalyer current as a function of the longitudinal position. The white lines are defined as in Fig.~\ref{fgr:IV_5.2nm_maps}. (c) Interlayer current along the device at $V_{\rm GS}=0.6$~V for different values of $L_{\rm OV}$.}
  \label{fgr:I-LOV_maps}
\end{figure*}

Next, we consider the impact of the length of the overlap region on the device operation. Fig.~\ref{fgr:IV_LOV} reports the transfer characteristics obtained for different values of $L_{\rm OV}$, while the interlayer current spectrum close to the ON and OFF states for $L_{\rm OV}=23.6$ nm is reported in Fig.~\ref{fgr:I-LOV_maps} (a) and (b), respectively.  

Fig.~\ref{fgr:I-LOV_maps} (a) shows that in the OFF state, the interlayer tunneling mostly occurs in narrow energy windows, corresponding to the thermionic and tunneling peaks of the longitudinal current. Spatially, the interlayer current follows an oscillating behavior, associated to the presence of quantum-well-like states. They result from the confinement enforced in the Au$_2$S by the ending of the monolayer on the source side and by the channel barrier on the other side.
The weak degradation trend of SS as $L_{\rm OV}$ increases, visible in Fig.~\ref{fgr:IV_LOV}, reflects a modest enhancement of the electron thermalization as the overlap region is enlarged, due to the increase of the number of electron-phonon scattering events.

%Phonons assist the process both by promoting electrons at higher energies in the source and by enabling the inelastic vertical tunneling from the top of the conduction band in the source. 
%the distance covered by the electrons increases.  
In the ON state (Fig.~\ref{fgr:I-LOV_maps} (b)), the interlayer tunneling principally occurs in the energy window between the bottom of the two bands and is mostly localized at the edges of the heterojunction. 
%[where the local maxima of DOS are located]. 
The spatial distribution of the interlayer current for different values of $L_{\rm OV}$ is plotted in Fig.~\ref{fgr:I-LOV_maps} (c). It can be noticed that, as $L_{\rm OV}$ decreases, the current concentrates more on the right edge of the heterojunction. Actually, the tunneling probability is particularly high at this edge and, for $L_{\rm OV}$ not too small ($\ge 5.9$ nm in Fig.~\ref{fgr:IV_LOV}), the increase of the current density at this point approximately compensates its decrease with $L_{\rm OV}$ in the rest of the overlap region. This explains the weak dependence of the current at large gate overdrives on $L_{\rm OV}$, noticeable in Fig.~\ref{fgr:IV_LOV}. Overall, the device appears weakly sensitive to $L_{\rm OV}$ in any operation regime. This represents a relevant property, as $L_{\rm OV}$ is a design parameter that can be difficult to finely control in practical manufacturing processes.

\begin{figure*}
    \centering
	\includegraphics[width=1.5\columnwidth]{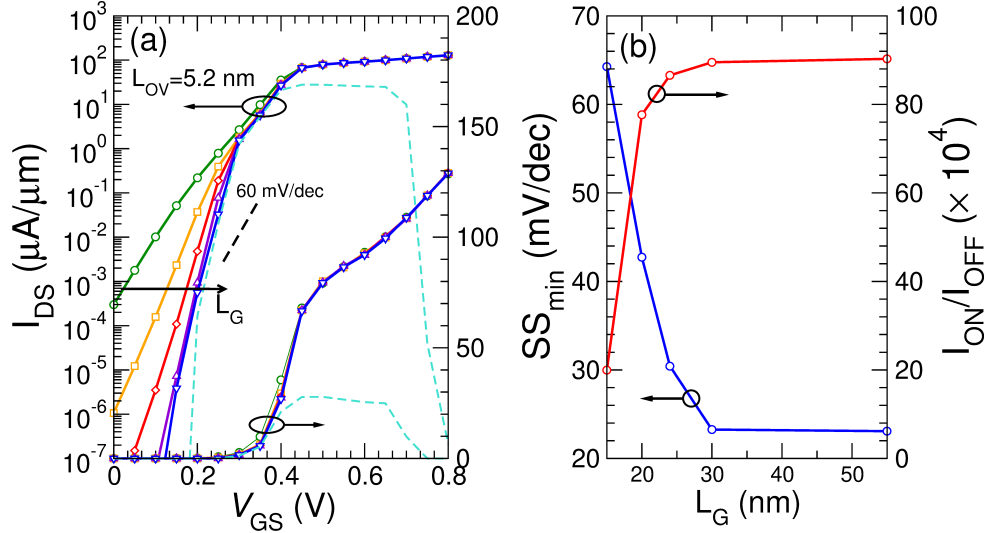}
	\caption{\protect \footnotesize	(a) Transfer characteristics of the transistor for $L_{\rm OV}=5.2$~nm and $L_{\rm G}=15$, 20, 25, 30 and 55~nm. $V_{DS}=0.4$~V. The dashed curves are obtained by taking into account only the elastic interaction between electrons and acoustic phonons. (b) minimum of SS for $I_{\rm DS}\ge I_{\rm OFF}=10^{-4}$~$\mu$A/$\mu$m and $I_{\rm ON}/I_{\rm OFF}$ ratio as a function of $L_{\rm G}$.}
  \label{fgr:I-LG}
\end{figure*}

To assess the scaling behavior of the device, we report in Fig.~\ref{fgr:I-LG} (a) the transfer characteristics for $L_{\rm OV}=5.2$~nm and different values of $L_{\rm G}$. The minimum value of SS over the $V_{\rm GS}$ window of interest and the $I_{\rm ON}/I_{\rm OFF}$ ratio are plotted as a function of $L_{\rm G}$ in Fig.~\ref{fgr:I-LG} (b). Scaling down the gate length only influences the subthreshold regime, and, particularly, degrades the subthreshold swing. To achieve a subthermionic behavior, the gate length must be larger than  15~nm, while values above 30~nm improve only negligibly the performance of the device. The SS value of $\approx 23$~mV/dec, obtained for $L_{\rm G}=55$ nm, represents a lower limit for this device, set by the intrinsic phonon absorption. The pronounced dependence of the current on the gate length is due to the small effective mass of Au$_2$S, which results in a rapid increase of the source-to-drain tunneling probability as the gate length is reduced.

In order to assess the role played by the strength of the electron-optical phonon coupling, we also investigated the limit of vanishing interactions between electrons and optical phonons in the case $L_{\rm G}=55$ nm. The corresponding transfer characteristic (dashed curves in Fig.~\ref{fgr:I-LG} (a)) is obtained by only including in the simulation the interactions between electrons and elastic acoustic phonons. According to the previous analysis, lower values of SS are achieved, particularly for $V_{\rm GS}\lesssim 0.2$~V, namely when the top of channel barrier is above the highest energy of the KTlCl$_4$ conduction band. Indeed, in these conditions the electrons injected by the source contact necessarily need to increase their energy by absorbing optical phonons in order to overcome the barrier. On the other hand, we notice that the current at large gate overdrives significantly decreases. The reason is that optical phonons enable further connections between high density of state regions, therefore activating new high-conductive paths. The consequent enhancement of the interlayer tunneling eventually dominates over the increase of the backscattering. The transfer characteristics with and without electron-optical phonon interactions almost overlie for intermediate $V_{\rm GS}$ values, for which the energy redistribution of electrons driven by optical phonons is weaker ({\it cf.} map (b) in Fig.~\ref{fgr:IV_5.2nm_maps}). The sudden drop of the current for $V_{\rm GS}>0.65$~V is an irrealistic artifact of the simulation, due to the strong backscattering of electrons elastically traveling in KTlCl$_4$ from the down-bending potential profile.

As the dashed curve in Fig.~\ref{fgr:I-LG} (a) suggests, in general our model predicts that sub-1~mV/dec SSs could be in principle attained by using channel materials with small enough electron-phonon scattering rates and large enough band gaps (to avoid interband tunneling between conduction and valence bands), and by properly extending the gate length in order to suppress the tunneling through the channel barrier. A favorable band alignment of the monolayers is furthermore needed to locate the smallest values of the SS in the $I_{\rm DS}$ window of interest for the operation of the device as a digital switch.

\begin{figure*}
    \centering
	\includegraphics[width=1.5\columnwidth]{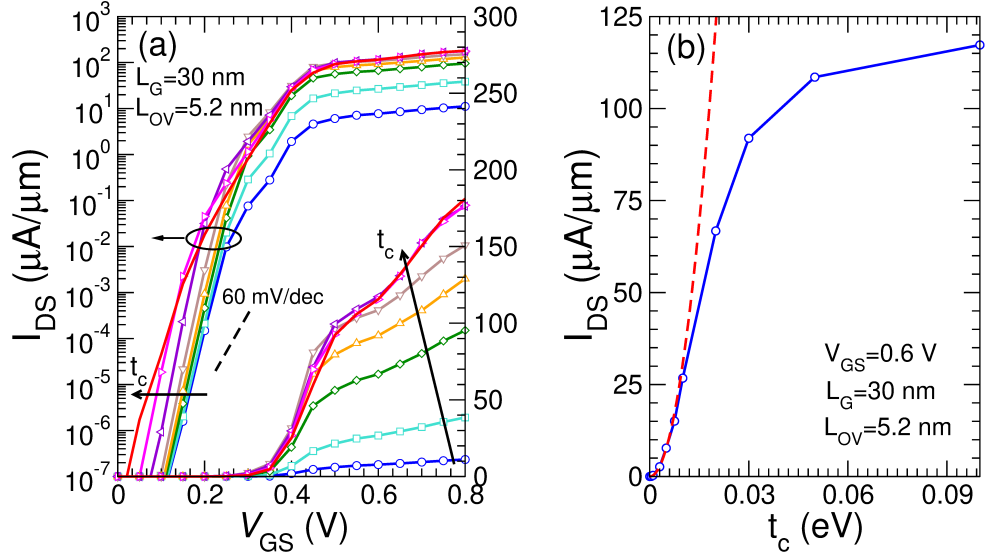}
	\caption{\protect \footnotesize	(a) Transfer characteristics of the transistor for $L_{\rm G}=30$~nm, $L_{\rm OV}=5.2$~nm and effective hopping values $t_c=5$, 10, 20, 30, 50, 100, 150 and 200~meV. $V_{DS}=0.4$~V. (b) $I_{\rm DS}$ at $V_{\rm GS}=0.6$~V as a function of $t_c$. The dashed curve is a parabolic fit for small values of  $t_c$.}
  \label{fgr:I-tc}
\end{figure*}

Finally, we explore the dependence of the device operation on the value of $t_{\rm c}$. We report in Fig.~\ref{fgr:I-tc} (a) the transfer characteristics of the device for $L_{\rm OV}=5.2$~nm, $L_{\rm G}=30$~nm and different values of $t_{\rm c}$. Different choices of $t_{\rm c}$ mainly impact the current at large gate overdrives, with a behavior that can be better understood from Fig.~\ref{fgr:I-tc} (b), where the current at $V_{\rm GS}=0.6$~V is plotted as a function of $t_{\rm c}$. For small values of the coupling, the current increases quadratically with $t_{\rm c}$, according to a first-order approximation of the interlayer current~\cite{Lannoo2004}, $I_{\rm int}\propto t_c^2\times D_{\rm T}\times D_{\rm B}$, where $D_{\rm T}$ and $D_{\rm B}$ are the density of states on the top and bottom layer. For $t_{\rm c}>100$~meV, the current saturates, indicating that it is not limited anymore by $t_{\rm c}$ itself, but by the low density of states of the Au$_2$S. This almost optimal regime is not far from that assumed in this paper by setting $t_{\rm c}=30$ meV. Our results should thus be considered close to an upper limit of the perfomance potentially attainable by the device.

%while it saturates for $t_{\rm c}>100$~meV. This can be easily explained on the basis of a first-order approximation of the interlayer current~\cite{Lannoo2004}, $I_{\rm int}\propto t_c^2\times D_{\rm T}\times D_{\rm B}$, where $D_{\rm T}$ and $D_{\rm B}$ are the density of states on the top and bottom layer. For small $t_{\rm c}$, the current is limited by $t_{\rm c}$ itself, while for large enough $t_{\rm c}$ the current is limited by the low density of states $D_{\rm T}$ of the Au$_2$S. This latter is essentially the case for the value of $t_{\rm c}=30$ meV chosen in this paper. As this represents an almost optimal situation, our results should be considered close to an upper limit of the perfomance potentially attainable by the device. 

\begin{figure}
    \centering
	\includegraphics[width=0.9\columnwidth]{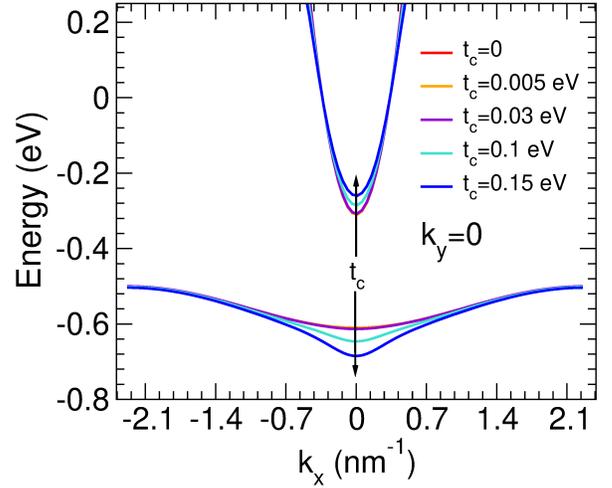}
	\caption{\protect \footnotesize	Cross-section at $k_y=0$ of the band structure of the KTlCl$_4$-Au$_2$S heterostructure for different values of $t_c$.}
  \label{fgr:BS_tc}
\end{figure}

The value of $t_{\rm c}$ also modulates the band alignment between the monolayers, as shown in Fig.~\ref{fgr:BS_tc}. The increasing separation of the bands as $t_{\rm c}$ increases results in the gradual shift of the transfer characteristics toward more negative $V_{\rm GS}$ visible in Fig.~\ref{fgr:I-tc} (a). On the other hand, it can be seen that the value of 30~meV for $t_{\rm c}$ is small enough to entails only slight modifications of the band structure of the coupled monolayers. It is thus compatible with weak interactions of van der Waals type. 

%We observe in passing that the actual value of $t_{\rm c}$ could in principle be determined by comparing %experimental measurements of the current as a function of the interlayer coupling with the theoretical %curve of Fig. ... The interlayer coupling can be in principle tuned 

\section{Conclusion}
We have proposed a new device concept for energy-efficient steep-slope FETs based on the combination of 2D materials with a weakly dispersive and a strongly dispersive band, interconnected through a van der Waals heterojunction. The low-dispersive material acts as the device source and provide an intrinsic cold-source energy-filtering effect on the electrons injected by the contact. The electron thermalization is by a large extent prevented by the low electron-phonon coupling in the strongly dispersive material implementing the channel and the drain of the device. Despite the presence of two materials, the device architecture reflects that of a standard MOSFET, in which the current is modulated by a thermionic barrier, instead of a tunnel barrier, as in tunnel FETs. Thus, according to previous studies, the proposed architecture is expected to be significantly less sensitive to the nonidealities that strongly degrade the performance of tunnel FETs, such as traps, band tails, and roughness. In order to demonstrate the device concept, we studied the operation and the performance of a device based on a KTlCl$_4$ and a $\alpha$-Au$_2$S monolayer. For a gate length of 30~nm and an equivalent gate oxide thickness of about 0.5~nm, this device can attain SSs close to 20~mV/dec and a current modulation of 6 orders of magnitude. On the other hand, it shows little sensitivity to the size of the overlap between the monolayers, a hardly controllable parameter in  manufacturing processes. The scaling properties of the device are dominated by the quite rapid increase of the source-to-drain tunneling as the gate length is decreased, favored by the smallness of the Au$_2$S effective mass. Finally, we remark that the occurrence of weakly dispersive, isolated bands around the Fermi energy is not uncommon in already reported 2D materials. We mention, for instance, the case of several monolayers ferromagnetic or antiferromagnetic in their ground state, such us CrBr$_2$, CrI$_2$, MnBr$_2$, MnCl$_2$, NiBr$_2$, NiCl$_2$, VOBr$_2$ and VOCl$_2$~\cite{mounet2018two}. Also, several 2D materials with remarkably small effective mass of electrons and holes, and sizable band gap, such as $\alpha$-AsSb~\cite{Zhao2017Giant} and GeSe~\cite{Hu2015GeSe}, have been proposed. Together with the conjectured abundance and variety of thermodynamically stable monolayers still to be explored~\cite{mounet2018two}, these considerations suggest that many opportunities to implement and optimize the proposed architecture may be found.

\begin{acknowledgments}
This work is partially supported by the European Commission through the Horizon 2020 FET Open Project QUEFORMAL (contract n. 829035) and by the Italian MIUR through the PRIN project FIVE2D.
J.C. acknowledges partial support from the Natural Science Foundation of Jiangsu Province under grant number BK20180456 and from the Key Laboratory for Information Science of Electromagnetic Waves (MoE) under grant number EMW201906.
\end{acknowledgments}

\appendix
\section{Computational Methods}

The DFT simulations were performed in a plane wave basis by means of the Quantum Espresso suite~\cite{Giannozzi2009,Giannozzi2017}. The KTlCl$_4$ monolayer was simulated by using scalar relativistic ultrasoft pseudopotentials based on the Perdew-Burke-Ernzerhof (PBE) exchange-correlation functional, while the Au$_2$S monolayer was simulated by using fully-relativistic norm-conserving pseudopotentials with a Heyd-Scuseria-Ernzerhof hybrid  functional. In both cases, a $12\times 12\times 1$ Monkhorst-Pack (M-P) k-point grid was used, with a cutoff energy of 50 Ry. Van der Waals interactions were taken into account through DFT-D3 Grimme's corrections~\cite{grimme2010consistent}.

Electron affinities were computed according to the definition, as the difference between the energy in the vacuum and the energy of the lowest unoccupied energy level in the materials.

The dielectric permittivity tensors were computed within the framework of the density functional perturbation theory (DFPT) as implemented in the ph.x code (included in the Quantum Espresso package). $12\times 12\times 1$ and $3\times 3\times 1$ M-P wave vector grids were used for electron and phonons, respectively and the energy cutoff was set at 60 Ry. The results were then interpolated on finer $80\times 80\times 1$ grids by means of the EPW code~\cite{PONCE2016}%\cite{Noffsinger2010EPW}. 
In order to extract from the simulation results, which include the effect of the vacuum layer needed to avoid interactions between the supercell replicas, the value of the dielectric tensor associated to the monolayers only, we resorted the principle of the series and parallel capacitances, as decribed in Ref.~\cite{laturia2018dielectric}. For the KTlCl$_4$, we found $\epsilon_{\parallel}=2.5$  and $\epsilon_{\perp}=2.9$, for the in-plane and out-of-plane dielectric permittivity, respectively. For the Au$_2$S, we found $\epsilon_{\parallel}=6.0$  and $\epsilon_{\perp}=9.1$.

The deformation potential of the dominant acoustic and optical phonon branches in KTlCl$_4$ were computed from the corresponding matrix elements. We found 0.25~eV and $2\times 10^8$~eV/cm, for the acoustic and the optical deformation potential, respectively. In the case of Au$_2$S, we used the acoustic deformation potential associated to the longitudinal modes provided in Ref.~\cite{wu2019two}. 
%We checked that the scattering with the transverse acoustic modes is negligible in comparison, as expected from the nearly spherical symmetry of the conduction band. 
The electron-polar optical phonon interaction in Au$_2$S was modeled through an effective deformation potential of $1.6\times 10^8$ eV/cm, computed by fitting the scattering rate of the electrons with the dominant polar-optical phonon mode. The interactions with flexural optical phonons are neglected, as they are expected to be strongly suppressed by the embodiment of the monolayers into the gate stacks. Moreover, in the case of Au$_2$S, the coupling of electrons with flexural phonons is zero to the first order, due to the existence of a glide symmetry plane~\cite{Fischetti2016Mermin}. Both the scattering rates and the electron-phonon matrix elements were computed  within the framework of the DFPT, by means of the codes ph.x and EPW.   

To the purpose of the simulation of transport, the device was modeled as translationally invariant in the transverse $y$ direction. Transport simulations were performed within the framework of the non-equilibrium Green's function formalism, by self-consistently solving the transport equations and the Poisson equation in the device cross-section. A set of 17 transverse wave vectors was used to sample the Brillouin zone of KTlCl$_4$, of Au$_2$S and of the KTlCl$_4$-Au$_2$S heterostructure. The electron-phonon interactions were described within the self-consistent Born approximation, by assuming an elastic and dispersionless approximation for acoustic and optical phonons, respectively.

\end{document}